\documentclass[fleqn,usenatbib]{rasti}
\usepackage{float}
\usepackage{algorithm}
\usepackage{algpseudocode}
\usepackage{graphicx}
\usepackage{amssymb}
\usepackage{amssymb,mathtools}
\usepackage{algpseudocode}
\usepackage{amsmath}
\usepackage{hyperref}
\usepackage{pslatex}
\usepackage{float}
\usepackage{xcolor}
\usepackage{multirow}
\usepackage{multicol}
\usepackage{bm}
\usepackage{siunitx}
\usepackage{lettrine}
\usepackage{makecell} 
\usepackage{newtxtext,newtxmath}

\usepackage[T1]{fontenc}

\DeclareRobustCommand{\VAN}[3]{#2}
\let\VANthebibliography\thebibliography
\def\thebibliography{\DeclareRobustCommand{\VAN}[3]{##3}\VANthebibliography}

\usepackage{graphicx}	
\usepackage{amsmath}	

\title{A Deep Neural Network Based Reverse Radio Spectrogram Search Algorithm}

\author[P.X Ma et al.]{
Peter Xiangyuan Ma,$^{1}$\thanks{E-mail: peterxy.ma@mail.utoronto.ca}
 Steve Croft,$^{2,3}$
 Chris Lintott$^{6}$,
Andrew P. V. Siemion$^{2,3,4,5}$
\\
$^{1}$Department of Mathematics, University of Toronto, 40 St.\ George Street, Toronto, ON M5S 2E4, Canada\\
$^{2}$Radio Astronomy Laboratory, 501 Campbell Hall, University of California, Berkeley, CA 94720, USA\\
$^{3}$SETI Institute, Mountain View, CA 94043, USA \\
$^{4}$Jodrell Bank Centre for Astrophysics, The University of Manchester, M13 9PL, UK \\
$^{5}$University of Malta, Institute of Space Sciences and Astronomy\\
$^{6}$Department of Physics, University of Oxford, Denys Wilkinson Building, Keble Road, Oxford, OX1 3RH, UK
}

\date{Accepted 07/12/2023. Received 05/12/2023; in original form 13/02/2023}

\pubyear{2023}

\begin{document}
\label{firstpage}
\pagerange{\pageref{firstpage}--\pageref{lastpage}}
\maketitle

\begin{abstract}

\textcolor{black}{Modern radio astronomy instruments generate vast amounts of data, and the increasingly challenging radio frequency interference (RFI) environment necessitates ever-more sophisticated RFI rejection algorithms. The ``needle in a haystack'' nature of searches for transients and technosignatures requires us to develop methods that can  determine whether a signal of interest has unique properties, or is a part of some larger set of pernicious RFI. In the past, this vetting has required onerous manual inspection of very large numbers of signals.} In this paper we present a fast and modular deep learning algorithm to search for lookalike signals of interest in radio spectrogram data. First, we trained a \textcolor{black}{\(\beta\)-Variational Autoencoder} on signals returned by an energy detection algorithm. We then adapted a positional embedding layer from classical Transformer architecture to a \textcolor{black}{embed additional metadata, which we demonstrate using a frequency-based embedding}. Next we used the encoder component of the \(\beta\)-Variational Autoencoder to extract features from small ($\sim 715$\,Hz, with a resolution of $2.79$\,Hz per frequency bin) windows in the radio spectrogram. We used our algorithm to conduct a search for a given query (encoded signal of interest) on a set of signals (encoded features of searched items) to produce the top candidates with similar features. We successfully demonstrate that the algorithm retrieves signals with similar appearance, given only the original radio spectrogram data. \textcolor{black}{This algorithm can be used to improve the efficiency of vetting signals of interest in technosignature searches, but could also be applied to a wider variety of searches for "lookalike" signals in large astronomical datasets.}
\end{abstract}


\section{Introduction}
\subsection{Interference Rejection in Radio Astronomy}
 
The rejection of radio frequency interference (RFI) is a perennial challenge for radio astronomy, particularly given the increase in satellite constellations that transmit at a range of radio frequencies and are detectable even at remote observing sites. RFI rejection is traditionally performed using statistical or machine learning techniques that risk rejecting a potential signal of interest (SOI) in the process of flagging RFI signals \citep{Pinchuk2022}. This is a particular problem when searching for astrophysical transients or for technosignatures, since the signal morphology may not be known ahead of time and it may be rejected before a human has the opportunity to review it. Novel deep learning algorithms have been successfully used to address these kinds of problems \citep{Ma2023}.

\textcolor{black}{Substantial work has been done to develop deep learning algorithms in radio astronomy for anomaly detection and RFI rejection. For example, \cite{Mesarcik2020} designed a novel approach to diagnose system health in modern radio telescopes using a convolutional variational autoencoder to embed high dimensional data into lower dimensional features, which guided telescope operators to better visually inspect failures and maintain system health. \cite{Mesarcik2022} proposed an alternative approach to RFI rejection by training models solely on uncontaminated emissions. This method then leverages Nearest-Latent-Neighbours (NLN) for novelty detection. Additionally, \cite{Mesarcik2023} developed a new framework for identifying both common and rare anomalies in radio telescopes through the use of self-supervised learning that combines context prediction and reconstruction losses. They were successful in categorizing the anomalies into 10 different classes such as lightning storms, or problematic amplifiers or antennas. \cite{Voncina} studied the use of deep learning based autoencoder models and independent component analysis (ICA), for anomaly detection in spectral data obtained from the LOFAR radio telescope. This work demonstrated the effectiveness of cascading convolutional autoencoder architecture in addressing their interference rejection problems.}

\textcolor{black}{Despite the work described above on grouping RFI or system faults into various classes, tools do not yet exist to help researchers contextualize individual signals of interest by searching for ``lookalikes'' on the basis of signal morphology.  In addition, past methods do not incorporate other scalar metadata. Our approach enables us to make explicit pairwise comparisons once a SOI has been received, in addition to enabling the inclusion of other data features.}

The importance of devising an algorithm to address this kind of problem arises in fields such as technosignature searches where the vetting of candidate signals (e.g.\ BLC1; \citealt{blc1}) is rather onerous using existing methods. \textcolor{black}{One of the current strategies employed is to manually search for all lookalike signals across multiple observations, flagging candidates as interference should they visually appear similar, along with other potential metrics in metadata.} The ability to find additional examples of a given SOI at different times or different parts of the observed band can be invaluable in determining the nature of the signal. The focus of this paper is to develop an algorithm that allows us to reverse search a given SOI and return similar signals.
\
 We begin by taking inspiration from well known reverse image search algorithms.
\subsection{Classical Reverse Image Search Algorithms}
\label{sec:classical}
Reverse image searches are commonly employed in search engines and social media. These algorithms take an input image as a search term and return the locations of the same or similar images from across the web. These sites use algorithms like Scale-Invariant Feature Transforms (SIFT; 
 \citealt{sift}), Maximally Stable Extremal Regions (MSER; \citealt{MSER}) or Bag of Words/Vocabulary Tree \citep{BoW} to power their search. \textcolor{black}{ What all these algorithms have in common are that they are, or make use of, feature extractors or embedding techniques. This means that they take spatial data (like images) and attempt to translate them into feature data (like vectors) while trying to capture large scale structure such as various spatial relationships present in the original input. Many approaches to computer vision problems in the past focused on trying to represent image-like data as vectors. This was motivated by the availability of a richer set of tools to deal with vector-like data (e.g., decision trees, or support vector machines) than was available for images. Today, the rise of deep learning has enabled algorithms like convolutional neural networks to learn directly on images without handcrafted labels. This approach is also applicable to radio spectrograms, which are two-dimensional image-like datasets.}
 
 \textcolor{black}{ In addition to the use of reverse image search in industry, similar ideas have also been explored in astronomy: for example the use of self-supervised similarity search on the Dark Energy Spectroscopic Instrument datasets \citep{desi} or the use of more traditional cluster algorithms like $S$-means clustering used for LIGO's data mining strategies for similarity searches \citep{ligo_x}. Researchers have also used contrastive learning, e.g., \cite{Stein2022} where use authors used this method for applications including galaxy morphology classification, strong lens discovery, and similarity searches.}

 Here we will briefly review some of the existing approaches used to search for patterns in images, and extend them to our work.

\subsubsection{Scale-Invariant Feature Transforms}
\label{sec:sift}
A Scale-Invariant Feature Transform (SIFT) is an algorithm that takes an image, locates special ``locally distinct points'', and describes the features around such points using measurements such as the gradient in intensity or maximum intensity, etc.\ \citep{sift}. This process produces pairs of vectors: the coordinates of the locally distinct point, and the corresponding descriptor vector that contains the features of that point.

The locally distinct points are produced from a process of difference of Gaussians performed at varying resolutions of the images. To compute the descriptor vector we measure the gradients in pixel intensity over a region about the locally distinct points. Together this builds a feature extractor. Finally, to perform a search, we match the key points between images and check the similarity between the descriptor vectors. More concretely we can use a technique called a Vocabulary Tree, described in section~\ref{sec:vocab}.

\subsubsection{Maximally stable extremal regions}
The maximally stable extremal regions (MSER) algorithm, like SIFT, attempts to find key points in the image. Specifically, it looks for objects called ``blobs'', defined as areas of an image that have connected elements, contrasting backgrounds, and close-to-uniform pixel intensities 
 \citep{MSER}.

MSER works by taking various thresholds in the range (0, 255) and blacking or whitening out pixels depending on this threshold. The blacking and whitening of pixels create the ``blobs''. We perform a connected component analysis and track how the blobs evolve as we adjust the threshold. MSER helps pull out small regions of images that contain distinctive boundaries which we can use to label distinct points and compute descriptor vectors for each blob. These can then be used to compare and match.

\subsubsection{Bag of Words / Vocabulary Tree}
\label{sec:vocab}
The Vocabulary Tree approach in computer vision effectively implements a Bag of Words (BoW) model to compare images to each other \citep{BoW}. We first run a local feature extractor like SIFT (section~\ref{sec:sift}). We then construct a ``codebook'' containing codewords that describe several similar patches identified by SIFT. One simple means of determining codewords is to use K-Means clustering to produce centroids, which we denote as codewords. We can then compare and contrast images by comparing the corresponding codebooks using a variety of algorithms. However, a key limitation in this approach is that SIFT ignores the spatial relationships between local patches, which we know is incredibly important in describing an image accurately. This leaves room for improvement.

\subsection{Deep Learning for Computer Vision}
 Deep learning overcomes some of the shortcomings of the aforementioned algorithms and has been proven to effectively solve a wide range of image problems with the advent of Convolutional Neural Networks (CNN; \citealt{cnn}). CNNs are simple in that they are a traditional neural network with the addition of convolutional layers. These layers operate by performing convolution operations between input data and a kernel. These operations have built-in inductive bias. For example, these operations are equivalent to translations that the algorithm exploits as fundamental properties in images. More concretely, this is because in images, small translations often do not change the prediction outcomes \citep{ML_textbook_Goodfellow}. For example, if we take a picture of a dog and move the dog 5 pixels to the right, the image is still that of a dog. By using a convolutional layer we are baking this assumption into the model without getting the model to learn it from scratch. More formally this allows us to restrict the priors on the weights thus reducing the size of the model and allowing vast improvements in both performance and scale. More specifically CNNs have been proven to be efficient feature extractors of both local and global features which address the shortcomings described in Section~\ref{sec:vocab}. Due to the exceptional performance of CNNs on a wide range of image tasks, they have been widely adopted in industry for a variety of computer vision challenges \citep{imagenet, industry} including for reverse image search algorithms \citep{reverse_img}.

\subsubsection{Autoencoders}
\label{sec:autoencoder_introduction}
Autoencoders (AE) are special cases of CNNs. \textcolor{black}{They are  unsupervised models that take input data and learn to compress them down to the most important features, similarly to the the feature extractor algorithms described in Section~\ref{sec:classical} with the difference being these algorithms learn what to extract from the dataset itself.} AE's achieve this by employing a symmetrical CNN architecture that takes in an input and outputs a reconstruction of the original input. However, these models have a bottleneck that constricts the flow of data \citep{ae}. This bottleneck divides the network into two: an encoder and a decoder. The encoder takes an input image and compresses the data through the bottleneck, and the decoder attempts to reverse that process. Intuitively this builds an efficient feature extractor since the encoder is attempting to reduce the input into the most important features such that the decoder can reconstruct the original input \citep{ML_textbook_Goodfellow}. This kind of automatic feature extraction technique will serve as the core of our approach.

\subsubsection{Variational Autoencoders}
\textcolor{black}{Variational Autoencoders \citep{vae} are an improvement on the traditional Autoencoder. The issue with traditional Autoencoders is that the encoded feature space is not smooth. This means that interpolating between encoded feature vectors can generate nonsensical output. We would like the encoded feature space to be smooth, meaning if we gradually perturb the encoded feature vector, once decoded, the output should also smoothly transition from the original. This allows practitioners to use these models as generators in creating new data. In our case this also allows us to construct a much more interpretable feature space. Perhaps even more importantly for astronomers, these interpretable features may relate the latent space to physical attributes of the data, which in many situations have smooth features. This allows us to make better comparisons when conducting a reverse search.}

\subsubsection{Interpretability}
\textcolor{black}{Despite the power of VAEs in extracting features from their inputs, these models, like many deep learning algorithms, can be something of a black box, making interpretation of results difficult. To combat this, algorithms like \(\beta\)-VAE \citep{bvae} have been used in radio astronomy to extract features from large datasets \citep{Ma2023}. These \(\beta\)-VAE use a scalar factor \(\beta\) to weigh the importance of minimizing the KL-Divergence loss term. This has the effect of biasing the training towards building a model whose encoded features are statistically independent from each encoded feature dimension. This means that individual entries in the encoded feature vectors correspond to a relevant feature in the original input. For example, if the input image was a white circle against a black background, then after encoding it using this model, we might perturb the first entry of the encoded feature, and see that the new decoded image shows a white circle translated to the left by a few pixels. This would then indicate that the first entry of the encoded feature vector extracted represents horizontal spatial translation, and spatial translation is a human-interpretable feature. This is an important consideration in building a reverse search algorithm as it allows us to search for specific morphological features in radio spectrograms.}

\subsection{Applications of Autoencoders in Astronomy}
\textcolor{black}{Autoencoders have become an indispensable tool in astronomy. These neural networks excel in many problem domains. For example, they help in denoising and enhancing astronomical images in noisy observations \citep{denoise} or cleaning noisy data in gravitational wave interferometers \citep{ligo_ae}. Moreover, they streamline the analysis of high-dimensional datasets by reducing dimensionality while preserving critical features, aiding in the identification of signals of interest and transient events \citep{Ma2023} and in aiding similarity searches, for example galaxy morphologies \citep{reverse_search_galaxy}. Autoencoders' anomaly detection capabilities facilitate the discovery of rare physical phenomena \citep{ae_anomaly}, offering a data-driven and automated approach for new astrophysical discoveries. They also contribute to data compression \citep{compress}, and automated object recognition \citep{galaxy_ae_class}, and  ultimately reducing human bias in data interpretation. In the evolving landscape of astronomy, these techniques are poised to play an increasingly pivotal role in how astronomers perform data analysis.}

\subsection{Problem Statement}
\textcolor{black}{To summarize the core problem we are trying to solve, }although industry-standard, out-of-the-box algorithms exist, they fail to satisfy our needs. Firstly, these traditional (not deep learning) techniques disregard large-scale spatial features, as described in section \ref{sec:vocab}. Traditional algorithms fail to capture complex structures within the data. This is partially the reason why traditional computer vision techniques are being superseded by deep learning in tackling computer vision problems. Secondly, under certain circumstances, we want to bias our model towards particular features, for example, frequency. Currently, there is no encoding of information on the radio frequency of the signal in any of these methods.

During our search process, sometimes we wish to match signals that not only appear similar to each other, but also are located in similar regions of the frequency band. On the surface, this appears to be a trivial problem: simply filter out the candidates after the search is completed. However, there exists a trade-off between visually similar signals and signals that are just close by in frequency. Thus if one wants to search for both similar appearing signals and signals close by in frequency one needs to ``bake in'' the information about the frequency into the search process rather than apply it as a secondary filter. Classical search algorithms do not have an out-of-the-box solution to embedding complex features such as signal frequency.

The central questions we want to answer are: \textit{Given a signal in a radio spectrogram, can we find all lookalike signals? \textcolor{black}{Can we also encode additional metadata not included the original spectrogram, such as the frequency position? }Can we make these approaches modular? \textcolor{black}{And can we do so in a way that is to some extent human interpretable?}}

\section{Methods}
We begin by outlining the structure of our approach. First, we build a feature extractor using an \textcolor{black}{\(\beta\)-VAE including a hyperparameter tuning section}. To do so we describe our data source in section \ref{sec:datasource}, the cleaning of the data in section \ref{sec:energy}, and data preprocessing in section 
\ref{sec:preprocessing}. We then move to build the model in 
\ref{sec:autoencoder}. We train and test the model in section \ref{sec:training}. We extend the model capabilities using a \textit{modular} frequency embedding strategy discussed in \ref{sec:emebedding}, and finally assemble the algorithm in section  
\ref{sec:search_algo}.

\subsection{Data Source}
\label{sec:datasource}
The training, validation, and testing dataset are derived from real observations from Breakthrough Listen's Green Bank Telescope $1.8-2.9$\,GHz dataset, sourced from several
different observational campaigns \citep{enriquez2017turbo}. We used the high-frequency resolution data product where each frequency bin is \(2.79\)\,Hz wide \citep{data}, since our goal is to tackle data that contains RFI, and RFI can have fine resolution in the frequency domain. The data are open source and are available from the Breakthrough Listen Open Data Archive\footnote{\url{http://seti.berkeley.edu/opendata}}.

\subsection{Data Filtering Energy Detection}
\label{sec:energy}
Since we are working with real data, we need to balance the training features. Large regions of the spectrograms consist of Gaussian noise, so if we did not filter the data to extract signals, then the resulting model would be biased towards generating noise rather than real signals. To perform this filtering we apply a simple energy detection process\footnote{\url{https://github.com/FX196/SETI-Energy-Detection}}.

First we perform a simple bandpass filtering on the entire observation. This removes the polyphase filterbank shape imposed by the first of a two-stage channelization procedure \citep{data}. This is done by collapsing the spectrogram in the time dimension and fitting a piece-wise polynomial spline to each $\sim 3$\,MHz-wide coarse channel. We subtract the fitted polynomial from the data. Finally, we iterate through windows of size 715\,Hz and search for excess energy above the expectations of Gaussian random noise. We chose an S-score threshold of 512 \citep{scipy_s}. 

We also perform a complementary energy detection process where we select regions consistent with Gaussian noise by inverting the threshold condition. When constructing the training set we use an equal number of ``signal'' and ``noise'' regions, thus balancing the dataset.

\subsection{Data and Preprocessing}
\label{sec:preprocessing}
To create the training set we randomly draw 30 observations (which are 5 minutes of telescope recording time at a particular band using the Greenbank Telescope) from a total of 12,000, and draw an additional six observations for the test set (excluding those already selected as part of the training set). We chose 30 and 6 to keep computing time reasonable. Running energy detection required 30 minutes to process each file and with 30 examples it already provided more than 2 million training samples which is more than enough to demonstrate our algorithm's capabilities. The spectrograms have a time resolution of  18.25\,s and a frequency resolution of 2.79\,Hz, giving the dataset \(3\times 10^{8}\) frequency channels. 

\begin{figure*}
  \centering
  \includegraphics[width=\linewidth]{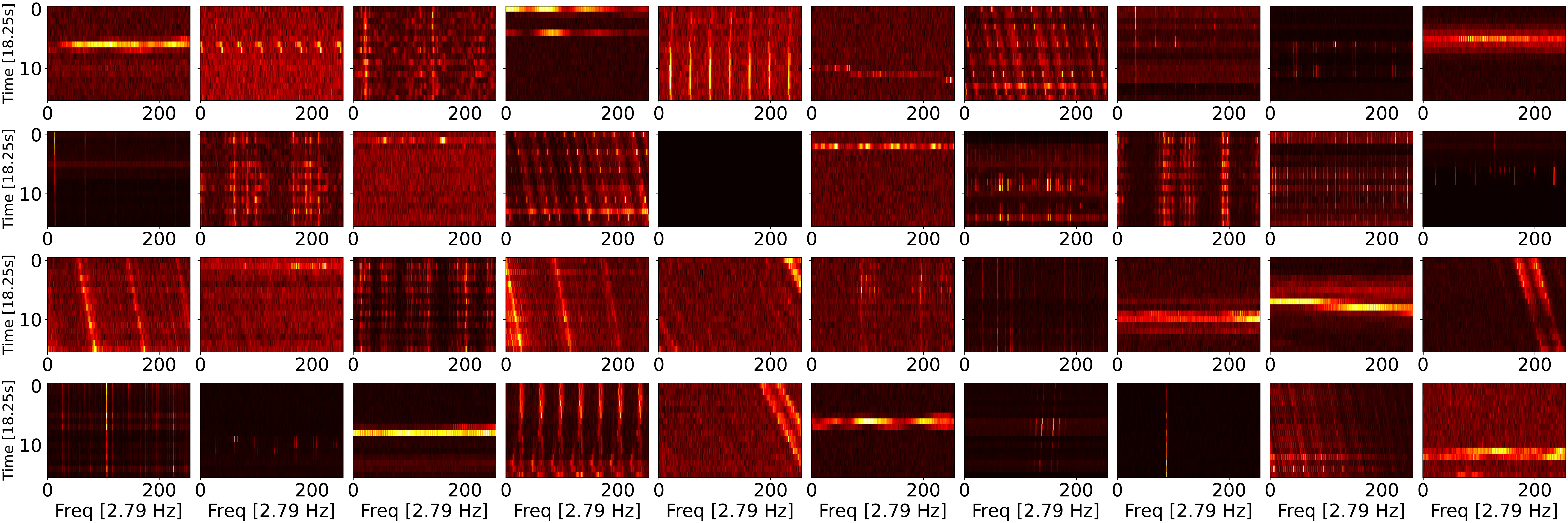}
  \caption{A random sample of training examples. A variety of RFI signals are seen in these snippets.}
  \label{fig:traing_example}
\end{figure*}

We then split the band into $\sim 715$\,Hz ``snippets''. This resulted in a training set of approximately 2 million snippets and a test set of 300,000 snippets. We then log normalize the data, add a constant to make all the values positive, and scale the data to have a final range within 0 and 1. Examples of the resulting snippets are shown in Figure~\ref{fig:traing_example}. Note that the normalization is done per snippet independently of each other sample. These snippets are used as inputs to the autoencoder, and the targets are the same snippets. 

\subsection{\textcolor{black}{\(\beta\)- Variational Autoencoder}}
\label{sec:autoencoder}
As described in section \ref{sec:autoencoder_introduction}, our autoencoder consists of an encoder and a decoder \citep{ae} \textcolor{black}{with input sizes of 16x256 pixels which represents bins in our spectrogram}. The encoder consists of five convolutional layers \citep{cnn}, with filter sizes of 16, 32, 32, 32, and 32 respectively. In between each layer is a 2-D Maxpool layer \citep{imagenet} of size (1,2). A batch normalization layer \citep{batchnorm} is included between the final maxpool layer and the convolutional layer. This is followed by three dense layers of size 32, 16, and 5 respectively. All the activations used are ReLu \citep{relu} activations. The model was built using Tensorflow \citep{tensorflow} and Keras \citep{chollet2015keras}.

\begin{figure*}
  \centering
  \includegraphics[width=\linewidth]{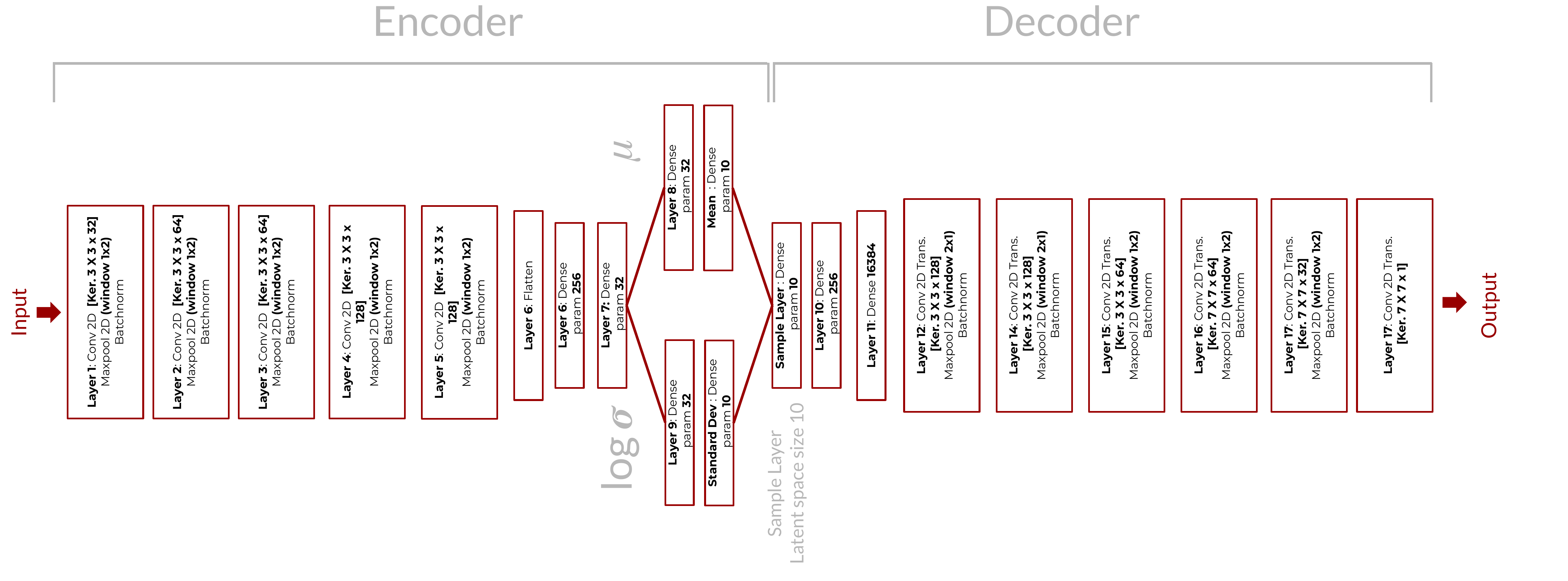}
  \caption{A representation of our VAE model. The encoder shows a progressive compression of data, forcing the network to decide the most important features of the original spectrogram.}
  \label{fig:MODEL}
\end{figure*}

The decoder is similar but in reverse order. We once again have dense layers of size 16 and 32, where the output is then reshaped and fed into a convolutional transpose \citep{ae} layer which upscales the image back to the same dimension of the input. Between the five convolutional layers, we have batch normalization and maxpool size of (2,1). The model architecture is shown in Figure~\ref{fig:MODEL}.

\textcolor{black}{In addition should one face a problem with varying input sizes of data, a popular method is to use Multiscale CNN instead of traditional CNN layers \citep{multiscalereview}. Some techniques used in these kinds of architectures have a pyramid or tower like structure. Similar to the SIFT algorithm, these CNNs typically have multiple branches or paths that process the input data at different resolutions or simply repeat operations but at different pixel resolutions and combining these features together \citep{multiscale}.}

\subsection{Training and Validation}
\label{sec:training}
\begin{figure*}
  \centering
  \includegraphics[width=\linewidth]{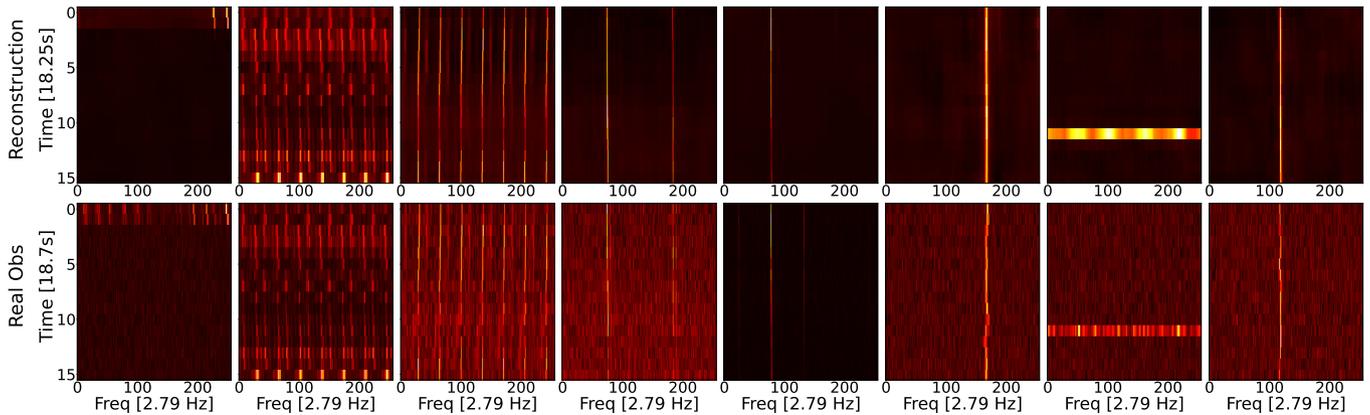}
  \caption{Eight randomly drawn real observations (top row) and their corresponding autoencoder reconstructions (bottom row). The autoencoder can reconstruct the signals in the input data. We see that the reconstruction of the signal appears good, whereas the reconstruction of the noise is slightly poorer. This isn't a concern since the main focus should be on the signal.}
  \label{fig:reconstruction_test}
\end{figure*}

The training scheme is standard. We fitted the model using the ADAM \citep{adam} optimizer with a learning rate of \(1\times 10^{-4}\), utilizing an early stopping routine with the patience of 10 epochs trained in batches of 16 samples, for 100 epochs. We then visually evaluated how well the model reconstructed the original spectrogram. Figure~\ref{fig:traing_example} shows a few randomly drawn examples. The reconstruction achieves our desired level of accuracy in that visually the reconstructions appear similar to the inputs. We proceed with this model (Figure~\ref{fig:reconstruction_test}).

\subsection{Embedding Additional Features}
\label{sec:emebedding}
\textcolor{black}{The approaches discussed above do not make use of  additional metadata that could be useful for inference, such as frequency, time, antenna characteristics, pointing direction, quality of data recorded, etc. Incorporating other scalar data modalities into our search is nontrivial as data may be discontinuous, discrete, or irregularly spaced. To implement this approach, we draw from research using positional embeddings used in modern Large Language Models (LLM's) \citep{llm_astro}.}
\textcolor{black}{As a proof of principle we demonstrate the effectiveness of embedding additional scalar data using frequency information.}

\subsubsection{Frequency Embedding}
Signal morphology varies with frequency, given that transmitters of different types occupy different regions of the band. For example, WiFi and blacktooth signals are common in our spectra around 2.4\,GHz. However, during our training phase and in the construction of our model this frequency information is not preserved in the feature extractor. Should a user choose to search for signals that are also similar in frequency this information would be lost. However, we can add frequency information to the feature vector.

Initially, this appears trivial --- one can simply extend the feature vector by a dimension and add the frequency information here. However, this would mean that frequency is an orthogonal feature to all the other extracted features. We expect, however, that signal morphology is correlated to some extent with frequency. We know this because certain transmitters have particular morphologies of which remain relatively close in frequency throughout observation times. To effectively exploit this additional information we borrow techniques from Natural Language Processing (specifically transformer architectures) to encode frequency information in feature vectors called Positional Embeddings \citep{attention}.

Positional embeddings work by taking encoded feature vectors and perturbing these vectors by small offsets based on their position to obtain an ideal balance between correct adjustment and over-adjustment which could lead to confusion with other signals. One can imagine this as trying to build miniature clusters in some high-dimensional feature space where each element of the cluster is unique but the elements of the cluster don't intersect with other clusters.

The adjustment vector is found using Equation~\ref{eq:embedding}, adapted from the paper ``Attention Is All You Need'' \citep{attention},
\begin{equation}
\label{eq:embedding}
P(k, i) = 
\begin{cases}
    \sin(\frac{k}{n^{\frac{i}{d}}}) , \quad i \text{ even}\\
    \cos(\frac{k}{n^\frac{i-1}{d}}), \quad i \text{ odd}\\
\end{cases}
\end{equation}

The variable \(k\) is the index in position, so if we have a sequence of length \(L\), we pick an integer \(k\in [0, L]\). The index in the feature vector is denoted \(i\), and \(i\in[0,d]\) where \(d\) is the dimension of the embedding space. The variable \(n\) is a tunable hyperparameter, where \citeauthor{attention} choose \(n = 10,000\). These equations ought to satisfy two conditions:
\begin{enumerate}
    \item The adjustments are unique for a given frequency index \(k\).
    \item The adjustments are bounded, in order to prevent the adjustment vector from over-adjusting.
\end{enumerate}

\begin{figure}
  \centering
  \includegraphics[width=1.2\linewidth]{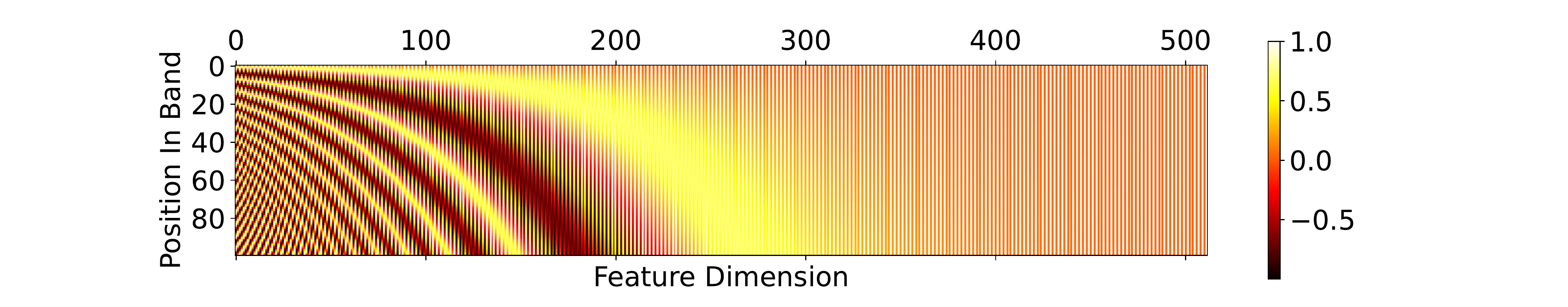}
  \caption{A visualization of the patterns in the embedding confirms that they are unique for each position we encode. We used a dimension of 512 and a sequence length of 100 for demonstration purposes; in the actual algorithm, we used a dimension of 4 and a sequence of 1000. }
  \label{fig:embeddings}
\end{figure}

The functions are all bounded between \([0,1]\) satisfying condition (ii). The sinusoidal functions are unique to each position, satisfying condition (i). We can measure similarity in positions using metrics such as cosine similarity or Euclidean distance. The embeddings are visualized in Figure~\ref{fig:embeddings}. 

The choice of sequence length determines how many frequency chunks the band is broken into. Our choice of  1000 bins (giving $\sim 1$\,MHz chunks) is a compromise between small chunks (which increase computational resource intensity) and large chunks (which reduce accuracy).

\subsection{\textcolor{black}{Hyperparameter Tuning}}
\textcolor{black}{We performed hyperparameter tuning in order to find the best configurations for our model. For our \(\beta\)-VAE model we search a space of 864 variations and each time train with 50 epochs. Using a Bayesian optimization technique, we sample a total of 128 variations out of all possible combinations of model variations. This hyperparameter tuning used a variation of the silhouette score (Section~\ref{sec:silhouette}) to determine how well it can cluster signals with similar morphologies. We redefined the intracluster score as the standard deviation of vectors within a given cluster, and the intercluster score as the distances between the center (mean position) of each cluster. These variations increase the speed of hyperparameter tuning while still measuring clustering loss. Parameters of the best performing models are highlighted in bold in Table~\ref{tab:4model_param}. Note that we tuned the \(\beta\) parameter manually, since \(\beta\) affects the disentanglement score, which is optimized differently to the silhouette score.} 

\begin{table}
    \centering
        \resizebox{1\linewidth}{!}{%
  { \color{black}\begin{tabular}{p{5cm}c}
  \Xhline{2\arrayrulewidth}
\multicolumn{2}{c}{Autoencoder}\\
\Xhline{2\arrayrulewidth}
Parameters &  Values\\

\Xhline{2\arrayrulewidth}
Latent Dimension & 3, 4, 5, 6, 7, ,8, \textbf{10}\\
First Convolution Kernel & \textbf{3}, 8, 16, 32\\
Second Convolution Kernel & 32, \textbf{64}\\
Third Convolution Kernel & \textbf{64}, 128\\
Dense Layer 1 & 64, 128, \textbf{256}\\
Dense Layer 2 & 16, 32, \textbf{64}\\
Learning Rate & 1e-3, \textbf{5e-4}, 1e-4 \\
\hline
\multicolumn{2}{c}{\(\beta\)-VAE } \\
\Xhline{2\arrayrulewidth}
Parameters &  Values\\
\hline
Latent Dimension & 3, 4, 5, 6, 7, ,8, \textbf{10}\\
First Convolution Kernel & 3, 8, 16, \textbf{32}\\
Second Convolution Kernel & 32, \textbf{64}\\
Third Convolution Kernel & 64, \textbf{128}\\
Dense Layer 1 & 64, 128, \textbf{256}\\
Dense Layer 2 & 16, \textbf{32}, 64\\
Learning Rate & \textbf{1e-3}, 5e-4, 1e-4\\
\hline

\hline
\end{tabular}}
}
\caption{ \textcolor{black}{Parameter values searched for each of the four sets of alternative models. The values in bold indicate the best performing model.}}
\label{tab:4model_param}
\end{table}
\subsection{The Search Algorithm}
\label{sec:search_algo}
\begin{figure}
  \centering
  \includegraphics[width=0.8\linewidth]{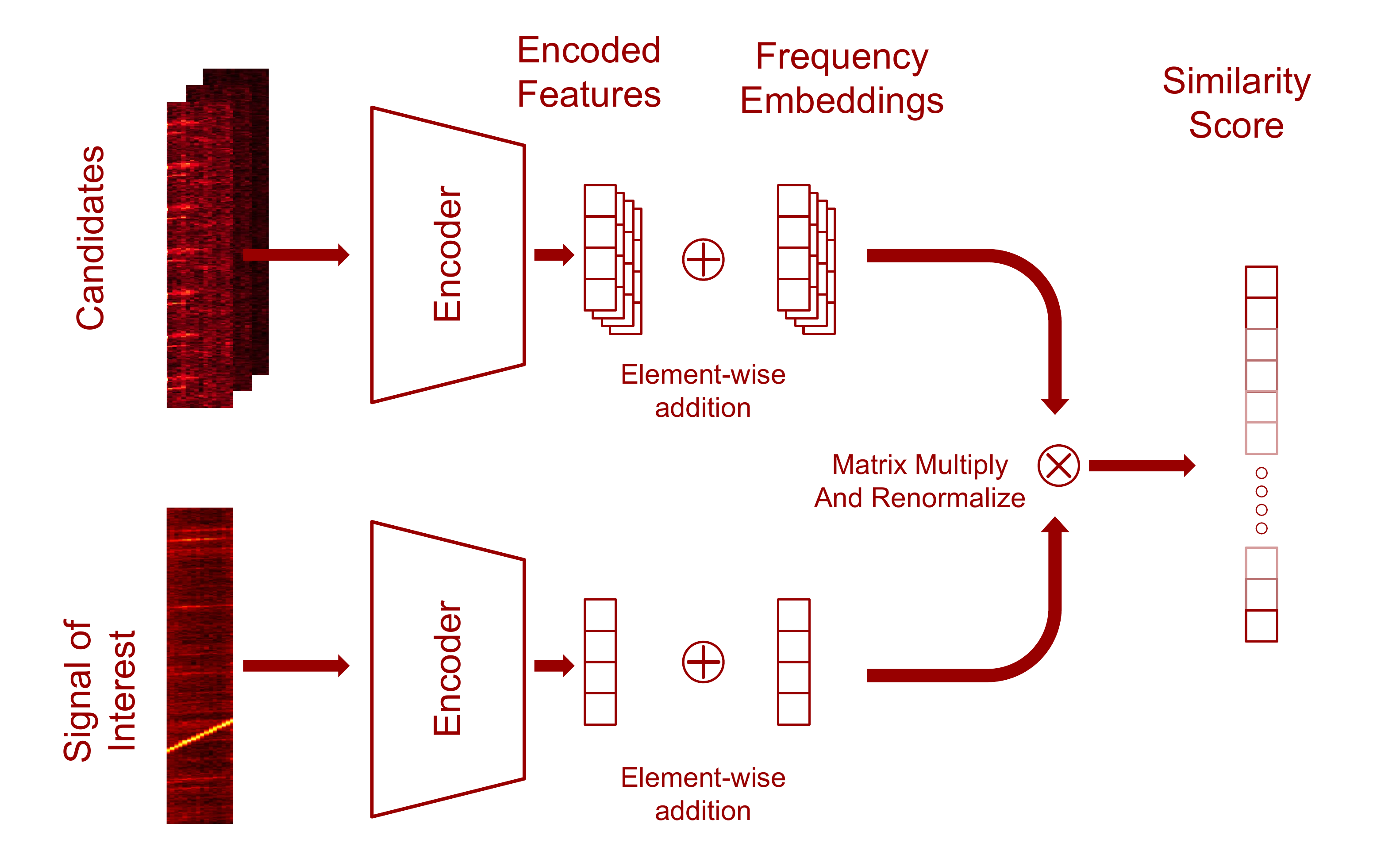}
  \caption{Visualization of the search process and the flow of data. We first extract features from the SOI and the set of possible candidates. Then we apply frequency embedding, and finally, we produce the similarity scores by matrix multiplication.  }
  \label{fig:SEARCH}
\end{figure}

We now combine the components of our algorithm and use them to perform a search. The search has three elements. First, the feature extractor, using the encoder of the Autoencoder, extracts features from the spectrogram. Second, we index and construct the frequency embedding for each of the extracted features and add those to the encoded vectors. We repeat this process with the SOI. Finally, we compute similarity scores between the SOI and the images in the search list. This similarity score is the cosine similarity metric and is computed efficiently by performing a matrix multiplication and renormalizing by the respective vector norms. The procedure is visualized in Figure~\ref{fig:SEARCH}.

\section{Results}
\subsection{\textcolor{black}{Testing Metrics and Results}}

\label{sec:metrics}
\subsubsection{\textcolor{black}{Silhouette Score}\label{sec:silhouette}}
\textcolor{black}{To compare our algorithm's capabilities to other deep learning techniques, we use a silhouette score \citep{silhouettes}. This silhouette score computes how well an algorithm is able to cluster data points in some high dimensional space. The score measures how effectively an algorithm can group similar features together while separating them from other clusters of similar features. The score ranges from $-1$ (poorly clustered) to 1 (well-clustered). }

\textcolor{black}{We constructed a test dataset by injecting simulated signals into real observations. We then constructed 100 classes, where each class has the same signal injected into a randomly sampled spectrogram from the test set, except their frequency position is varied. In each class there are 1000 samples. When we search for lookalike signals we want to find matches in drift rate, signal-to-noise ratio (SNR), and signal width. The signal frequency is not important in searching for lookalike signals.}

\textcolor{black}{We ran the above simulation 10 times, generating new signals by uniformly sampling from a range of SNR (20 to 70), drift rates ($-2$ to 2\,Hz/s ) and signal width (20 to 70\,Hz). These test ranges were chosen to correspond to signals that are easily found on visual inspection (e.g bright enough to be seen, but no so bright so as to overwhelm other background signals in the spectrograms). We record the mean and standard deviation of the silhouette scores for each set of 10 simulations. The final testing dataset of real observations contained 50,000 examples sampled randomly from a total test set of 300,000 examples. }

\textcolor{black}{We benchmark our algorithm against other approaches. In the first approach, we apply no additional preprocessing to the input and simply take the image data and unravel the spectrogram as a big vector. By unravel we are referring to iteratively taking the columns or rows and appending them into one long vector. The second approach uses the traditional BoW model plus SIFT algorithm to extract features. The third approach uses ResNet-50 \citep{resnet}, removing its final layer which is used for class predictions. Fourthly, we train a simple autoencoder from scratch on the same dataset as our \(\beta\)-VAE model. Lastly we test our \(\beta\)-VAE model. The results are shown in Table~\ref{tab:scores}.}

\begin{table}
    \centering
        \resizebox{0.6\linewidth}{!}{%
    \begin{tabular}{cc}
\hline
Model & \textcolor{black}{Silhouette Score} \\
\hline
Naive Model  & $-0.37\pm 0.06$\\
BoW/ SIFT  & $-0.34 \pm 0.03$\\
ResNet-50  & $-0.19 \pm 0.05$\\
Autoencoder & $-0.22 \pm 0.01$\\
$\mathbf{\beta}\textbf{-VAE}$ & $\mathbf{-0.12} \pm \mathbf{0.01}$\\

\hline
\end{tabular}
}
\caption{Silhouette scores of each model averaged over 10 separate tests. Scores closer to +1 represent improved clustering. The larger the better and the best results are in bold.}
\label{tab:scores}
\end{table}
\textcolor{black}{We observe that the \(\beta\)-VAE model produced better results. We interpret these scores to align with our understanding of each model. We see that in all cases the values are negative; we believe this is in part a result of our generative parameters. In creating the test set we used smooth and continuous transformations of an injected signal. 
This makes it challenging to separate signals into discrete clusters (producing a high inter-cluster score), resulting in negative Silhouette scores. However despite this the \(\beta\)-VAE model still outperforms other approaches as it constructs a latent space that better represents the generative parameters of the test dataset due to the model's disentanglement property. However to supplement this score we investigate a separate metric in section \ref{sec:clustering}.}

\subsubsection{Cluster Metric}
\label{sec:clustering}
\textcolor{black}{To further compare our approach to other algorithms we can investigate how tightly clustered classes of signals are.  We quantify this tightness by finding the maximum euclidean distance between the centre feature vector and every feature vector with the same label (i.e., finding the edge of the cluster) and then scaling this maximum by the mean of all the distances so that the scale is comparable. The centre feature vector is obtained by taking the average of all feature vectors of that label. This metric is answering the question of ``given a set of signals with similar features, how similar do these features appear in the latent space?''. Results are presented in Table~\ref{tab:cluster_score}. }
\begin{table}
    \centering
        \resizebox{0.6\linewidth}{!}{%
    \begin{tabular}{cc}
\hline
Model & \textcolor{black}{Clustering Metric} \\
\hline
Naive Model  & $129.03\pm 7.88$\\
BoW/ SIFT  & $3.09\pm 0.13$\\
ResNet-50  & $3.59 \pm 0.12$\\
Autoencoder & $2.67 \pm 0.11$\\
$\mathbf{\beta}\textbf{-VAE}$ & $\mathbf{2.16} \pm \mathbf{0.06}$\\
\hline
\end{tabular}
}
\caption{These clustering scores of each model are averaged over 10 separate tests. The smaller the scores the better and the best score is in bold.}
\label{tab:cluster_score}
\end{table}

\textcolor{black}{We note that our scores are consistent with our understanding of our latent spaces. Firstly, the $\beta$ -VAE model produces a smaller clustering metric since the latent space is continuous and thus smoothly changing one feature (in this case the frequency position of the signal) should yield small changes in distance in the latent space. In contrast other models use no additional representation learning techniques, resulting in a less well behaved feature space, and thus larger clustering metrics. We surmise that the BoW model produces smaller scores due to the reduced dimensionality of the data, and is therefore less susceptible to the curse of dimensionality  \citep{curse}. }

\subsubsection{\textcolor{black}{Disentanglement Score}}
\textcolor{black}{In order to evaluate how well our models were able to learn disentangled features, which are both indications of interpretability and independence of the encoded features, we implement a disentanglement metric as described by \cite{bvae}.}

\textcolor{black}{To understand how this score measures manages to capture both independence and interpretablility of the latent features, an effective approach would involve devising a data representation where a model encodes these features into \textit{separate latent components} such that a simple linear classifier can identify a signal's generative parameters (SNR, drift rate, frequency width, frequency position). For instance, if a linear classifier can determine the correct fixed generative parameter based on the encoded features when varying all other generative parameters, this indicates that there exists a linear or ``obvious'' relationship between the input data and generative parameter. For example, if we held the drift rate constant, a linear model would still be able to identify that the class of ``drift rate'' generative factors are held constant even when other parameters like the SNR or the frequency positions are varied.
Thus the higher generalization accuracy of the linear model, the better. Note that we only test the deep learning models using this metric, in order to better understand their inner workings and the interpretability of their features of the input data.}

\begin{table}
    \centering
        \resizebox{.7\linewidth}{!}{%
    \begin{tabular}{cc}
\hline
Model & \textcolor{black}{Disentanglement Score} \\
\hline
$\mathbf{\beta}\textbf{-VAE}$ & $\textbf{0.88} \pm \textbf{0.04}$\\
Autoencoder & $0.81 \pm 0.06$\\
ResNet-50  & $0.66 \pm 0.07$\\
\hline
\end{tabular}
}
\caption{Testing the Disentanglement Score of each model averaged over 10 separate tests totalling to 400,000 samples, with the uncertainty being the standard deviation of these separate tests. The larger value the better.}
\label{tab:scores}
\end{table}

\subsection{Visual Inspections}
\label{sec:visual}
Finally we test a variety of algorithms by using them to perform a reverse image search and visually inspecting the output. First, we apply the classical SIFT + BoW algorithm to the problem discussed in Section~\ref{sec:classical}. Second, we apply our algorithm \textcolor{black}{ using the \(\beta\) - VAE encoder} to the problem but without the frequency embedding step. Lastly, we apply our algorithm with frequency embeddings to the problem and assess the results, in order to demonstrate the modularity of our design. \textcolor{black}{Note that we decided to omit running similar visual comparisons to the other models, namely ResNet-50, Autoencoder and the Naive model, as those have been demonstrated to be less effective in grouping similar signals. In this section we focus on comparing the status quo to our new algorithm.}

To visually assess the ability of the algorithm in addition to the tests we have already demonstrated, we select filtered Energy Detection snippets at random and perform a search using the same observation set as used for the SOI observations.

\begin{figure*}
  \centering
  \includegraphics[width=0.8\linewidth]{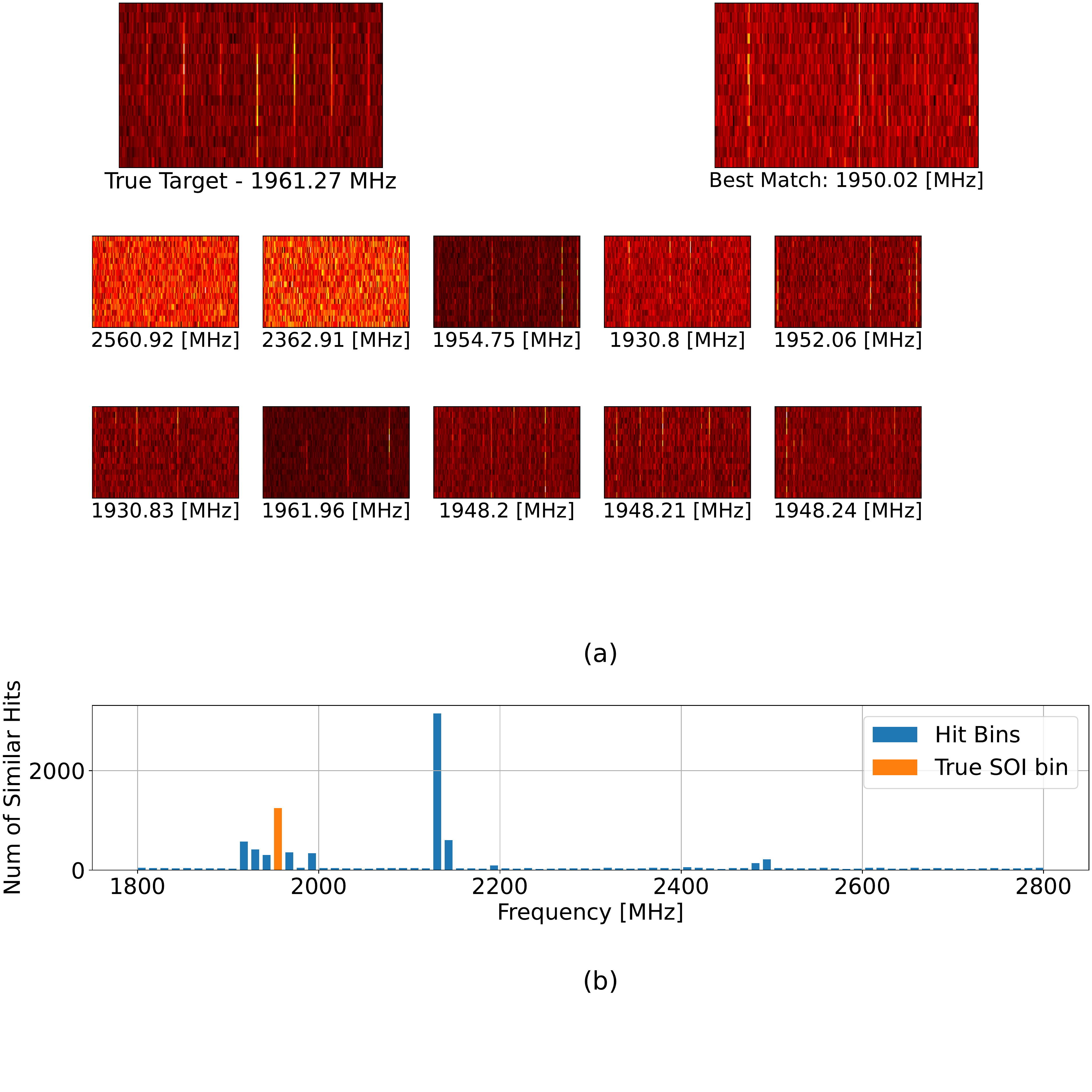}
  \caption{(a) The top 10 candidates with the closest similarity to the target [shown as the top left box], using the SIFT + BoW algorithm. The best candidate is shown in the top right box. Note: that this method did not find itself because this uses a KNN matching algorithm that excludes the trivial case. We also see that the algorithm tends to match noise with noise which is an indication of a poor feature extractor (b) Frequency distribution of the top 10,000 most similar hits. We see that the distribution does not align with where the true signal lies.}
  \label{fig:classical-search}
\end{figure*}

In the first method, we use the classical SIFT + Bag of Words algorithm described in section \ref{sec:classical}. We run SIFT through each window of the spectrogram and generate a set of descriptor vectors for each image. Then we use K-Means clustering with 800 centroids to build a codebook. The codebook is built by taking the fitted centroids as the codewords and then looping back in the descriptor vectors and assigning them a codeword. Then for each image, we created a histogram of codewords, hence treating it as a BoW model. Finally, we performed a match between the histogram of the SOI and the histograms of each candidate signal. The matches use the K-Nearest Neighbours approach \citep{KNN}. Results are shown in Figure~\ref{fig:classical-search}. \textcolor{black}{Firstly, we note that this approach fails to capture global structures in the input image. This is evident by the fact that the algorithm has matched noise with other noisy data which we know is an indication of potential flaws in the algorithm.} One downside to this approach is that the algorithm is very slow. For each observation, we need to generate millions of descriptor vectors and repeat the SIFT operation millions of times. Unlike deep learning-based approaches, these feature vectors can be generated in parallel on the GPU using much more efficient methods.

\begin{figure*}
  \centering
  \includegraphics[width=0.8\linewidth]{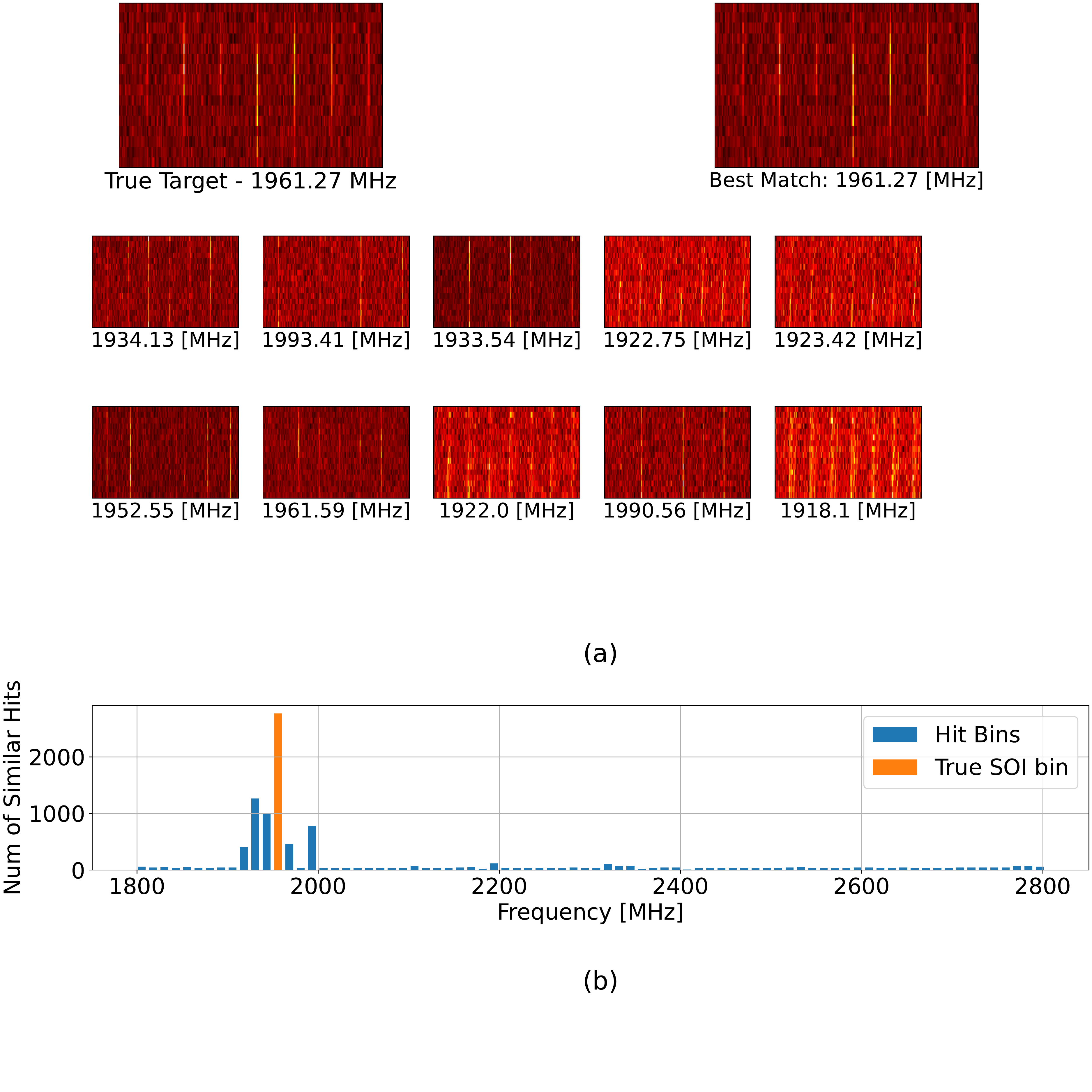}
  \caption{ (a) The top 10 candidates with the closest similarity to the target [shown as the top left box] for our deep learning algorithm with no frequency embedding. The best match is unsurprisingly, the input signal [top right box]. (b) Frequency distribution of the top 10,000 most similar hits.}
  \label{fig:full_search-noembedding}
\end{figure*}

Next, we applied our algorithm \textbf{\textit{without}} using frequency embedding techniques (Figure~\ref{fig:full_search-noembedding}). We see that this approach was able to retrieve more convincing candidates than the classical SIFT+ BoW approach. When we investigate \textit{where} the candidates lie in frequency space, we see that many are close to the frequency of the true signal (Figure~\ref{fig:full_search-noembedding}). \textcolor{black}{We see that the top search candidates visually appear more convincing than those retrieved by the classical algorithms. We also note that the retrieved signals tend to be found at similar frequencies to the input signal. Since many classes of transmitter emit only in their allocated spectral bands, this indicates that the algorithm has successfully identified different kinds of transmitters given the morphological characteristics of their transmissions alone.} 

\begin{figure*}
  \centering
  \includegraphics[width=0.8\linewidth]{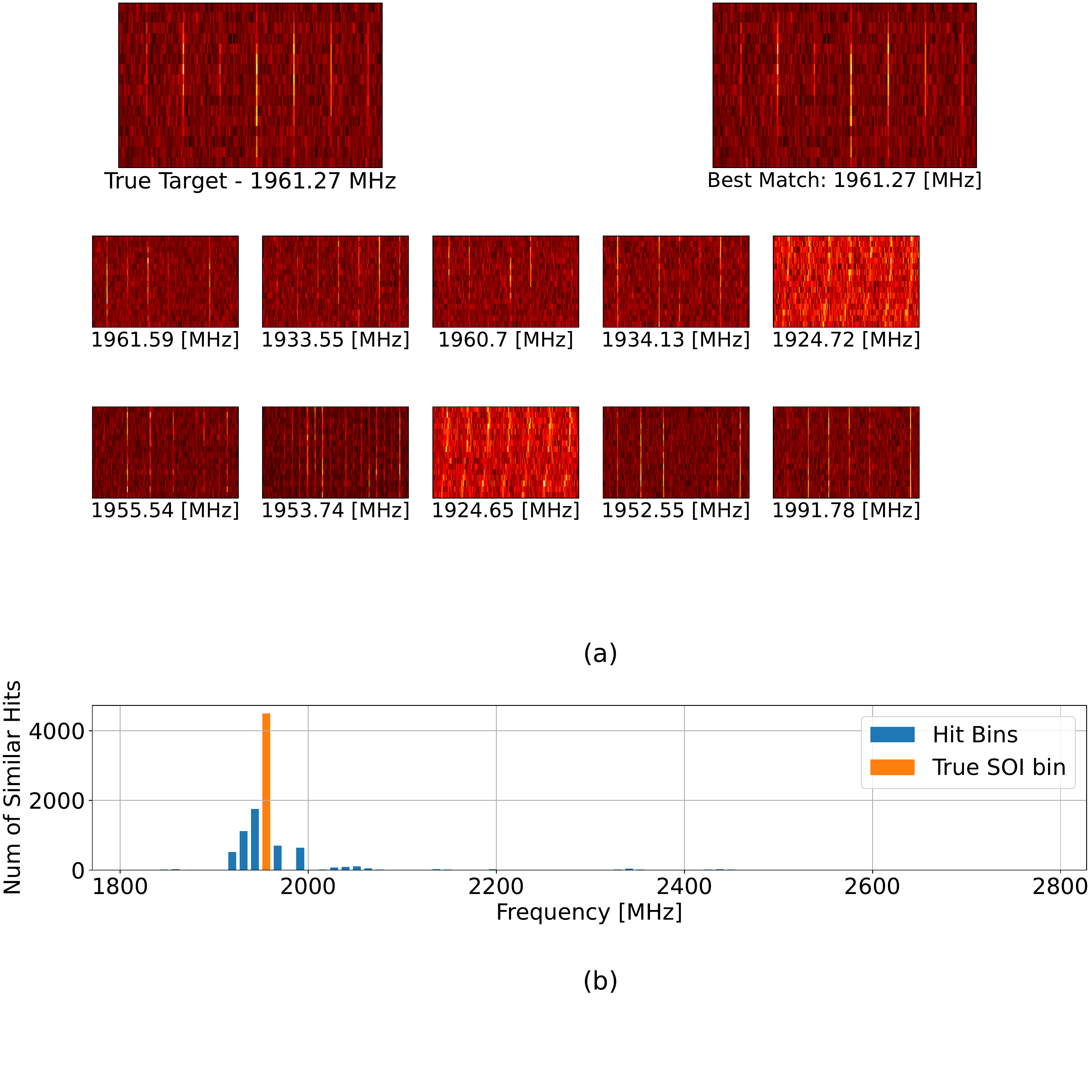}
  \caption{(a) The top 10 candidates with the closest similarity to the target [shown as the top left box] for our deep learning algorithm with frequency embedding. Again, the best match is the input signal [top right box]. (b) Frequency distribution of the top 10,000 most similar hits. The algorithm retrieves matches at frequencies close to the input frequency while maintaining similar signal morphology to the input.}
  \label{fig:full_search}
\end{figure*}

Finally, we applied our algorithm \textbf{\textit{with}} frequency embedding (Figure~\ref{fig:full_search}). The resulting signals show even higher visual similarity to the input signal. Additionally, the frequency distribution of the matches is closer to the frequency of the input signal, showing that the frequency embedding was successful. \textcolor{black}{The algorithm appears to produce even better morphological matches than the model without frequency embedding, and the frequencies of the output signals agree more closely with the input signal frequency. }

\textcolor{black}{In summary, the \(\beta\)-VAE algorithm outperforms other approaches both on visual inspection, as described in this section, and in the various metrics described in section \ref{sec:metrics}.}

\section{Discussion}
Our algorithm is more successful than classical approaches in returning similar candidates. The algorithm also runs approximately 20 times faster than classical implementations at our data center, \textcolor{black}{although this is partly because it can} take advantage of GPUs with deep learning hardware acceleration such as tensor cores. \textcolor{black}{The time taken to train deep learning approaches must also be taken into account when comparing efficiency, although training only needs to occur once, unless there are significant changes in the RFI environment at the telescope. In terms of scale current training time takes on the order of hours at our data center with modest compute.} Traditional matching techniques \textcolor{black}{also} suffer an inability to generalize large-scale features as discussed in Section~\ref{sec:classical}, and demonstrated in section \ref{sec:visual} -- specifically Figure~\ref{fig:classical-search} -- where the SIFT/ BoW algorithm fails. The addition of frequency embedding results in candidates that have a better match in frequency to the input signal, which is desirable when attempting to identify RFI from a particular class of transmitter. SIFT-based algorithms tend to match local features and thus have a tendency to match noise (see Figure~\ref{fig:classical-search}), unlike the deep learning based approaches where models were able to learn more global features resulting in better matches overall.

Looking forward we believe this algorithm has a wide range of practical use in transient radio astronomy. One particularly exciting use case is to help automate signal verification steps for candidate technosignature signals such as BLC1 \citep{blc1}. By automatically searching for potential lookalikes based on signal morphology, rather than relying on more low-level parameters of a signal, such as drift rate, signal width, or signal-to-noise ratio, we can do a better job of distinguishing between true anomalies, and RFI.

More generally, this technique can be used to build an RFI database. Many RFI databases consist simply of frequency ranges that are excluded (for example, the frequency ranges corresponding to GPS satellites). Our algorithm enables a finer-grained approach, comparing signal morphologies without necessitating the exclusion of large ranges in frequency, thus helping to make more efficient use of the spectrum and expanding the power of the search.

Another use case is to perform template searches. For example, one can build theoretical models that simulate some desired signal, and search the recorded observations for signals that match. 

The next step in the development of our algorithm is to modify it to handle a wider range of data products. For example, developing a model to handle arbitrary \textit{collections} of spectrogram data (data from different receivers or even telescopes).  Or perhaps building a model that can deal with cases where the spectral resolution varies  (e.g.\ combining data products with 3\,Hz resolution and 3\,kHz resolution; \citealt{data}).

We also plan to extend our approach to series (or cadences) of multiple observations, which intersperse scans of the target star with comparison scans of neighboring targets. Signals such as BLC1, which appear only in scans of the primary target, are consistent with being spatially localized on the sky. However, the dataset in which BLC1 was discovered also contained ``lookalike'' signals at other frequencies, indicating that they were likely due to particularly pernicious RFI. By applying our new methodology to cadences of data, we can much more easily locate lookalike signals at other frequencies, or in scans of other targets, providing an additional powerful means of screening technosignature candidates.

\section{Data Release and Code Availability}
The source code\footnote{\url{https://github.com/PetchMa/Reverse\_Radio\_Search}} and data\footnote{\url{http://seti.berkeley.edu/opendata}} are publicly available.

\section*{Acknowledgements}

Breakthrough Listen is managed by the Breakthrough Initiatives, sponsored by the Breakthrough Prize Foundation\footnote{\url{http://www.breakthroughinitiatives.org}}. We are grateful to the staff of the Green Bank Observatory for their help with installation and commissioning of the Breakthrough Listen backend instrument and extensive support during Breakthrough Listen observations. CJL thanks the Alfred P.~Sloan Foundation for support. We thank Yuhong Chen for his helpful discussion on the Energy Detection Algorithm.

\bibliographystyle{rasti}
\bibliography{ref.bib}


\label{lastpage}
\end{document}